\documentclass[cameraready]{Interspeech}

\title{Towards Privacy-Preserving ASR: Speaker-Level Machine Unlearning}

\author[affiliation={1}] {Seaone}{Ok}
\author[affiliation={2}]{Seungu}{Han}
\author[affiliation={1}]{Eungbeom}{Kim}
\author[affiliation={1,2,3}]{Kyogu}{Lee}

\address{
    $^1$ Interdisciplinary Program in Artificial Intelligence, Seoul National University \\
    $^2$ Department of Intelligence and Information, Seoul National University \\
    $^3$ Artificial Intelligence Institute, Seoul National University, Republic of Korea
}

\email{seaoneok@snu.ac.kr, hansw0326@snu.ac.kr, eb.kim@snu.ac.kr, kglee@snu.ac.kr}


\usepackage{comment}
\usepackage{graphicx}
\usepackage{xcolor}
\usepackage{tabularray}
\usepackage{booktabs}
\usepackage{multirow}
\usepackage[table]{xcolor}

\begin{document}

\maketitle

\begin{abstract}
Automatic Speech Recognition (ASR) systems are commonly designed to transcribe linguistic content independently of speaker identity. However, speaker-specific information can persist within the latent representations of ASR models, raising privacy concerns when training data includes sensitive individuals. While machine unlearning has been studied in several speech tasks, applying speaker unlearning to ASR remains challenging due to the strong entanglement of linguistic and speaker-related information in speech representations. In this work, we conduct a layer-wise analysis of Self-Supervised Learning (SSL) based ASR representations, revealing that speaker identity is primarily encoded in intermediate encoder layers. Based on this observation, we explore a feature-level unlearning strategy that targets these identity structures. Experiments on the VCTK corpus show that the proposed approach removes the influence of specific speakers while maintaining robust transcription performance.

\end{abstract}

\section{Introduction}
Machine unlearning has recently emerged as critical research frontier, driven by the need to comply with regulatory mandates such as the Right To Be Forgotten (RTBF) under the General Data Protection Regulation (GDPR)~\cite{rtbf}. These frameworks grant individuals the right to request that their data be removed from trained systems, posing a significant challenge when a user withdraws consent after a model has been deployed. 
Unlearning has expanded into the speech domain, particularly for tasks such as Keyword Spotting, Speaker Identification, and Text-to-Speech~\cite{speech_unlearning, slu_unlearning, tts_unlearning}. In these applications, the information targeted for unlearning are typically entangled with the primary goal of the model. However, unlearning remains underexplored in Automatic Speech Recognition (ASR), which is one of the foundational task of speech processing. As ASR fundamentally relies on broad generalization to transcribe effectively, it is commonly assumed that unlearning specific data could severely degrade its general transcription performance. 

Although ASR is typically formulated as a speaker-independent task, empirical evidence shows that modern ASR models still encode speaker-specific traits during training~\cite{asr_memory}.
This effect becomes pronounced in modern ASR pipelines, where large pretrained speech models are fine-tuned on relatively small, application-specific datasets. While such approaches enable strong performance in data-constrained scenarios like clinical or low-resource applications~\cite{clinicaltelephony, dutchdsr, parkinsondsr, low-resourcelang}, neural networks are known to memorize distinctive details of their training data beyond what is required for generalization~\cite{secretsharer, unintendedmem}. In context of ASR, this results in residual acoustic fingerprints that persist within the model representations. These signals can potentially be exploited through Membership Inference Attacks (MIAs)~\cite{mia, ssl_mia}, allowing an adversary to determine whether a particular speaker was included in the training data~\cite{asr_loss_mia}. Such risks raise serious privacy concerns~\cite{machineunlearning}, particularly when the training corpus involves sensitive cohorts, such as patients in clinical studies, where a privacy breach could lead to the re-identification of protected health information. 

Despite these risks, to the best of our knowledge, achieving speaker-level unlearning in ASR while preserving transcription performance has not yet been explored~\cite{orthograd}. This difficulty arises from the coupling of linguistic and paralinguistic information within speech representations. An effective unlearning strategy must therefore selectively remove speaker-specific traces while maintaining the general recognition performance, ensuring that speech from those speakers is transcribed with the same fidelity expected for any previously unseen speaker. 

In this work, we investigate speaker-level unlearning in ASR. Building on prior observations that speaker-related information tends to be localized within intermediate layers of speech encoders~\cite{wavlm}, we explore a representation-level intervention that targets these identity structures directly in the latent space. To enable controlled manipulation of speaker representations, we attach an auxiliary representation head and perform feature-level unlearning on the resulting embeddings. As one concrete approach, we introduce a simple dispersion objective that pushes representations of the forget speakers away from their latent centroids. Our experiments demonstrate that such targeted intervention can effectively removes speaker-specific traces while maintaining robust ASR performance. By achieving a favorable privacy-utility trade-off compared to existing unlearning baselines, our work provides a practical step toward privacy-preserving ASR systems.

\begin{figure*}[t]
    \centering
    \includegraphics[width=\textwidth]{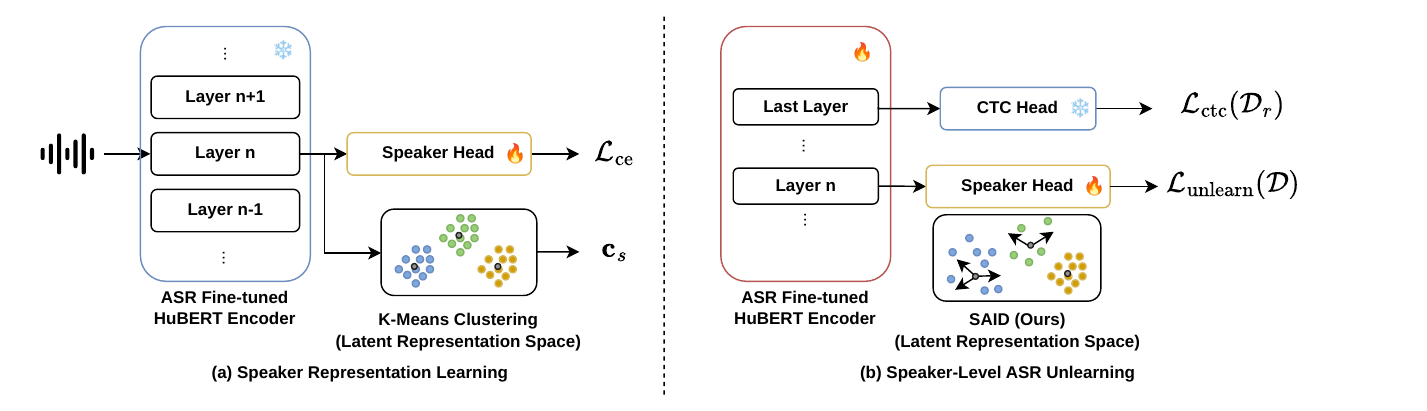}
    \caption{Overall framework of the proposed speaker-level ASR unlearning. \textbf{(a) Speaker Representation Learning} identifies speaker-specific distributions within the latent space of the ASR-tuned HuBERT-base model. K-Means clustering is employed to identify their geometric centroids $\mathbf{c}_s$. \textbf{(b) Speaker-Level ASR Unlearning} selectively dissolve speaker identity under each objective loss while maintaining transcription utility. Our mechanism applies a dispersion force to push latent embeddings from corresponding centroids, defined in the previous stage.}
    \label{fig:model_architecure}
    \vspace{-0.5cm}
\end{figure*}

\section{Related Work}
\subsection{Privacy Risks and Membership Inference in Speech Models}
Membership Inference Attacks reveal that training data can leave measurable traces in deployed models~\cite{mia, asr_loss_mia, audio_auditor}, including modern end-to-end ASR systems~\cite{vulnerable_asr_mia}. 
Moreover, SSL encoders such as HuBERT~\cite{hubert} have been shown to retain training-specific signals within their latent representations~\cite{ssl_mia}, suggesting that memorization may originate at the representation level and propagate to downstream tasks. While prior studies highlight the privacy risks of speech models, they primarily focus on auditing vulnerabilities rather than enabling mechanisms for identity erasure from trained ASR systems.




\subsection{Machine Unlearning}
Machine unlearning~\cite{machineunlearning} seeks to remove the influence of specific training data without retraining a model from scratch~\cite{mu_survey}. Existing approaches typically rely on parameter updates, such as gradient ascent or label corruption applied to the forget set~\cite{ga_rl}. More advanced methods introduce mechanisms such as knowledge distillation or structural model editing~\cite{unrolling_sgd, salun, kga, bad_t, mu_nlp}. Recent works has explored representation-level approaches that manipulate latent embeddings to control data influence more locally~\cite{duck}. 

Machine unlearning has been studied in several speech tasks~\cite{speech_unlearning, slu_unlearning, tts_unlearning}. However, applying unlearning to ASR remains particularly challenging due to the strong entanglement of linguistic and speaker-related information in speech representations. Consequently, only a limited number of studies have explored unlearning in ASR, including sample unlearning on LLM-based ASR architectures~\cite{llm_asr_unlearning} and optimization constraints during training~\cite{orthograd}, that suppress transcription performance for targeted speakers. However, this objective differs from the goal of general ASR systems, which are designed to operate independently of speaker identity. Existing studies have not yet demonstrated how to remove speaker-specific traces from ASR models while preserving their ability to transcribe speech for those speakers. 


\section{Method}
\subsection{Problem Definition}
\label{sec:problem definition}
We define the task of speaker-level machine unlearning for ASR. Let $\mathcal{D}= \{(x_i, y_i, s_i)\}^N_{i=1}$ denote the original training dataset, where $x_i$ is an input utterance, $y_i$ is its transcription $y_i$, and $s_i \in \mathcal{S}$ represents the speaker identity. Following the standard unlearning setup, we split $\mathcal{D}$ into a forget set $\mathcal{D}_f=\{(x,y,s)|s \in \mathcal{S}_f\}$, specified by a set of forget speakers $\mathcal{S}_f$, and a retain set $\mathcal{D}_r = \mathcal{D} \backslash \mathcal{D}_f$. Given a model $f_\theta$ trained on $\mathcal{D}$, the objective is to update its parameters to $\theta'$ such that $f_{\theta'}$ behaves similar to a gold model $f_{\theta*}$ trained exclusively on $\mathcal{D}_r$, without requiring full retraining. A successful unlearned model must satisfy the following three objectives.
\begin{itemize}
    \item \textbf{Utility preservation} The model must maintain its transcription performance on $\mathcal{D}_r$, ensuring that the linguistic mapping remains intact. 
    \item \textbf{Targeted Identity Erasure} Speaker-specific information associated with $\mathcal{S}_f$ must be removed to defend against membership inference attacks, while identity information of $\mathcal{S}_r$ remains unaffected.
    \item \textbf{Generalization} The model should maintain robust ASR performance on both forgotten and unseen speakers, treating forgotten identities as non-member data.
\end{itemize}

\subsection{Auxiliary Speaker Representation Head}
\label{sec:model architecture}
The overall architecture of ASR model consists of a pre-trained HuBERT-base encoder followed by a Connectionist Temporal Classifcation (CTC)~\cite{ctc} decoder. To obtain a controllable representation of speaker-related information, we attach an auxiliary speaker classification head to the hidden states of the 6th encoder layer. This choice follows prior analyses showing that speaker information tends to emerge in the early-to-middle layers of SSL speech encoders~\cite{wavlm}.
As illustrated in Figure~\ref{fig:model_architecure}(a), we first train this speaker head while freezing the encoder. This step learns a speaker embedding space $\mathbf{z}$ that captures identity-related structure without modifying the linguistic representations used by the ASR decoder. The subsequent unlearning stage is operated on this embedding space to enable representation-level interventions while minimizing disruption to transcription performance. 

\subsection{Representation-level unlearning objective}
The overall objective is defined as
\begin{align}
\mathcal{L}_{total} = \mathcal{L}_{\text{ctc}}(\mathcal{D}_r) + \lambda \mathcal{L}_{\text{unlearn}}
\end{align}
where the CTC loss is computed only on the retain data to maintain transcription performance. We exclude the forget set from the CTC objective, treating it as non-member data and relying on the inherent generalization ability of the model to transcribe forgotten speakers without explicitly optimizing on their samples during unlearning. Here, $\lambda$ balances the CTC loss and the unlearning objective; we set $\lambda=1$ in all experiments. To allow consistent comparison across different unlearning strategies, $\mathcal{L}_{\text{unlearn}}$ is applied directly to the speaker embedding $\mathbf{z}$. %
We use a unified formulation:
\begin{align}
\mathcal{L}_{\text{unlearn}} = \alpha\mathcal{L}_{\text{unlearn}}(\mathcal{D}_r) + \beta \mathcal{L}_{\text{unlearn}}(\mathcal{D}_f)
\end{align}
Here, $\mathcal{L}_{\text{unlearn}}(\mathcal{D}_f)$ induces forgetting by disrupting the identity representation of the target speakers, while $\mathcal{L}_{\text{unlearn}}(\mathcal{D}_r)$ stabilizes the representations of the retain speakers. In all experiments, we set $\alpha=1$ and $\beta=1$ for unlearning strategies. 

\subsection{Speaker-Aware Identity Dispersion}
We instantiate $\mathcal{L}_{\text{unlearn}}(\mathcal{D}_f)$ with a two-stage identity dispersion procedure that targets geometric structure of speaker clusters in the embedding space. 

\noindent \textbf{Stage 1: Centroid-based Distribution Mapping} Using the speaker embedding space defined in Section~\ref{sec:model architecture}, we estimate geometric anchors for the forget speakers. Although speaker labels are available, we intentionally apply K-Means clustering to the embeddings of the forget set to identify latent centroids $\textbf{c}_s$ corresponding to each speaker distribution. Speaker identity typically manifests as compact clusters in the representation space. Estimating the centroid of each cluster therefore provides a natural geometric anchor that summarizes identity-specific structure and serve as a reference points for targeted removal during unlearning. 

\noindent \textbf{Stage 2: Latent Identity Dispersion}
Once the centroids are estimated, we unfreeze the encoder and optimize the model to push the embeddings of forget samples away from their corresponding centroid. For a forget utterance $x_j$ belonging to speaker $s$, we apply a margin-based dispersion loss between the current embedding $\Psi_\theta(x_j)$ and its centroid anchor $\textbf{c}_s$. We refer to this objective as Speech-Aware Identity Dispersion (SAID). The SAID loss is defined as:
\begin{equation}
\mathcal{L}_{\text{SAID}} = \text{softplus} \left( \frac{\cos(\Psi_\theta(x_j), \mathbf{c}_s) - m}{\tau} \right)
\end{equation}
where $m$ is a distance margin and $\tau$ is a temperature scaling factor. Minimizing this objective progressively repels forget embeddings from their identity anchors, dispersing speaker-specific clusters in the latent space while preserving the broader linguistic representation required for ASR transcription.

\section{Experiments}
\subsection{Experimental Setup}
\noindent \textbf{Datasets} We evaluate our method on the VCTK corpus~\cite{vctk}, a multi-speaker English dataset of high-quality read speech, to simulate a realistic unlearning scenario where an ASR model is fine-tuned to a new domain.
We use the mic2 recordings and exclude p280 and p315 due to reported technical artifacts, resulting in a total of 108 speakers. All recordings are resampled to 16kHz. 
To rigorously evaluate speaker-level forgetting, we construct speaker-independent splits such that speaker identities are strictly disjoint across training and evaluation sets. 
The speakers are randomly partitioned into a retain set $\mathcal{D}_r$ of 95 speakers, a forget set $\mathcal{D}_f$ of 5 speakers and a test set $\mathcal{D}_t$ of 8 speakers. To mitigate potential speaker imbalance, we maintain a balanced distribution of gender and accent across the splits. To further distinguish between the removal of specific training samples and the erasure of speaker identity, both $\mathcal{D}_r$ and $\mathcal{D}_f$ are divided into training and evaluation subsets with a 9:1 ratio.

\noindent \textbf{Implementation Details} We first prepare the reference ASR models by fine-tuning a HuBERT-base encoder with the CTC loss. The Original Model $\mathcal{M}_{\text{orig}}$ is trained on $\mathcal{D}_{r, \text{train}} \cup \mathcal{D}_{f, \text{train}}$, while the Gold Model $\mathcal{M}_{\text{gold}}$ is trained exclusively on $\mathcal{D}_{r, \text{train}}$, simulating retraining without exposure to the forget data. 
To extract and manipulate speaker-specific information, we attach a speaker classification head to the hidden states of the 6th encoder layer. The speaker head is trained using a Cross-Entropy loss for 15 epochs with the learning rate of $3 \times10^{-4}$ on $\mathcal{D}_{r, \text{train}} \cup \mathcal{D}_{f, \text{train}}$, while the ASR encoder remains frozen to preserve the pre-trained representation space.
During the unlearning phase, this configuration is kept identical across all methods to ensure a unified representation space for fair comparison.
All unlearning methods, including baselines, are optimized using the same training settings with a learning rate of $1 \times10^{-5}$, and their objectives are applied to the speaker embedding $\mathbf{z}$ produced by the auxiliary speaker head. For SAID, the margin is set to $m=0$. All models are trained for 25 epochs using the AdamW~\cite{adamw} optimizer with a weight decay of 0.1.

\begin{table}[h]
\centering
\caption{Performance evaluation on $D_r$, $D_f$ and $D_t$. MIA scores $D_f$($\downarrow$) and $D_r$($\uparrow$) indicate the privacy protection levels for each unlearning method. Values closest to the gold standard are bolded, while the second-best results are underlined.}
\label{table:main_result}
\resizebox{0.95\linewidth}{!}{%
\begin{tabular}{l c c c c c}
\toprule
\multirow{2}{*}{\textbf{Methods}} & \multicolumn{3}{c}{\textbf{WER}} & \multicolumn{2}{c}{\textbf{MIA}} \\
\cmidrule(lr){2-4} \cmidrule(lr){5-6} & $D_r$ & $D_f$ & $D_t$ & $D_f$ & $D_r$ \\
\midrule
\rowcolor{blue!10} Original & 7.75 & 8.90 & 7.69 & 60.1 & 70.4 \\
\rowcolor{yellow!10} Gold & 7.43 & 11.05 & 7.06 & 52.7 & 69.8 \\
\midrule
Fine-Tuning       & 6.64 & 8.58 & 6.36 & 60.5 & \textbf{70.8} \\
Gradient Ascent~\cite{ga_rl} & 6.91 & 99.61 & 6.57 & 64.4 & 78.0 \\
Random Label~\cite{ga_rl}    & 6.91 & 7.60 & 6.64 & 47.6 & 75.1 \\
Bad-T~\cite{bad_t}           & 6.98 & 7.86 & 6.47 & \underline{53.4} & \underline{74.2} \\
SCRUB~\cite{scrub}           & 6.79 & 99.56 & 6.33 & 43.0 & 77.6 \\
DUCK~\cite{duck}            & \underline{7.07} & \underline{8.90} & \textbf{6.96} & 58.9 & 74.9 \\
\midrule
\textbf{SAID (Ours)} & \textbf{7.60} & \textbf{9.55} & \underline{7.30} & \textbf{52.6} & 74.8 \\
\bottomrule
\end{tabular}%
}
\end{table}

\subsection{Evaluation Metrics}
We evaluate the proposed unlearning method on the three criteria defined in Section~\ref{sec:problem definition}. 

\noindent \textbf{Performance and generalization} To measure the impact of unlearning on speech recognition performance, we report the Word Error Rate (WER) across multiple data partitions. For utility preservation, we evaluate WER on $\mathcal{D}_{r, \text{test}}$, which consists unseen utterances from retain speakers. This verifies that the model maintains its transcription capability for identities that should remain unaffected by the unlearning process. To assess generalization, we additionally report WER on $\mathcal{D}_{f, \text{test}}$ and $\mathcal{D}_t$. After unlearning, $\mathcal{D}_{f, \text{test}}$ is treated as an unseen speaker set. Thus, robust performance on $\mathcal{D}_{f, \text{test}}$ indicates that forgotten speakers are handled similarly to ordinary unseen speakers rather than suffering unintended degradation in recognition quality. Performance on $\mathcal{D}_t$ further reflects robustness to completely unseen identities. 

\noindent \textbf{Membership Inference Analysis} To quantify targeted identity erasure at the representation level, we conduct a representation-level MIA, following prior work~\cite{ssl_mia, ga_rl}. We train an external binary classifier as an attacker to distinguish between training members and non-members using pooled hidden representations extracted from the encoder. Specifically, the attacker is implemented as a logistic regression classifier operating on the encoder representations. This simple linear classifier serves as a probing model to measure whether membership signals remain linearly separable in the latent space. The classifier outputs a membership probability, indicating the confidence that a given sample belongs to the training data. In our experimental setting, the gold model represents ideal unlearning under the same data distribution. Effective unlearning is therefore characterized by the convergence toward the membership behavior of the gold model.

To provide a structured analysis of unlearning behavior, we evaluate membership inference performance across the following complementary partitions. 
\begin{itemize}
    \item \textbf{Speaker-level Forgetting} evaluates membership inference on unseen utterances from forget speakers $\mathcal{D}_{f, \text{test}}$. The absence of direct exposure to these samples during training isolates speaker-level identity traces rather than memorization of specific utterances. Successful unlearning should prevent these utterances from being recognized as members.
     \item \textbf{Retain Identity Preservation} evaluates whether unlearning inadvertently affects retain speakers $\mathcal{D}_{r, \text{test}}$. Membership separability for retain speakers should remain comparable to the gold model, confirming that the erasure process is targeted on the forget set and does not introduce unintended degradation. 
\end{itemize}

\section{Results}

\subsection{Main Results}
Table~\ref{table:main_result} shows the performance of the proposed method and various unlearning baselines in terms of ASR performance (WER) and privacy protection (MIA). These results highlight how different methods balance the three objectives defined in Section ~\ref{sec:problem definition}.  
Regarding targeted identity erasure, SAID achieves an MIA score of $52.6\%$ on $\mathcal{D}_{f,\text{test}}$, which is nearly identical to the privacy level of $\mathcal{M}_{\text{gold}}$. Since MIA was evaluated on $\mathcal{D}_{f, \text{test}}$, composed of unseen utterances from forgotten speakers, this result indicates that SAID successfully dissolves underlying speaker-specific acoustic features rather than merely erasing memorized training samples. In terms of utility, SAID maintains robust performance on $\mathcal{D}_{r,\text{test}}$ while showing a targeted increase in WER on $\mathcal{D}_f$ to closely align with $\mathcal{M}_{\text{gold}}$. This suggests a more authentic forgetting process compared to baselines that show lower WERs on $\mathcal{D}_f$ but exhibit higher privacy risks. Furthermore, SAID satisfies generalization on the unseen contexts, as evidenced by its improved performance on the unseen test set $\mathcal{D}_t$ compared to $\mathcal{M}_{\text{orig}}$. While some baselines achieve even lower WERs on $\mathcal{D}_t$, their suboptimal MIA scores suggest a failure to meet the primary privacy requirements of unlearning. These results position SAID as a more effective and reliable solution for speaker-level machine unlearning.
 
\begin{figure}[t]
    \centering
    \includegraphics[width=0.95\linewidth]{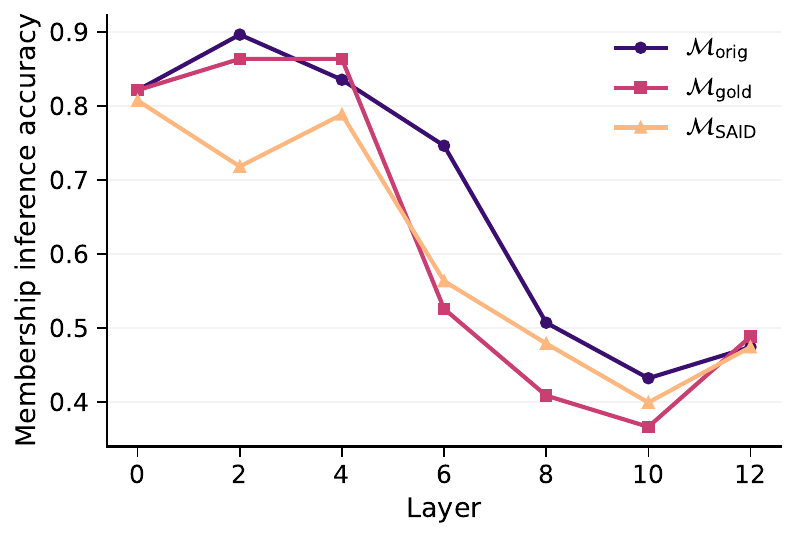}
    \caption{Membership inference accuracy evaluated on $\mathcal{D}_{f, \text{test}}$ for $\mathcal{M}_{\text{orig}}$ and $\mathcal{M}_{\text{gold}}$ across different HuBERT layers.}
    \label{fig:layer_probing}
    \vspace{-0.5cm}
\end{figure}

\subsection{Analysis}
\noindent\textbf{Layer-wise Analysis for Identity Localization} To identify the optimal layer for speaker identity manipulation, we conduct a layer-wise analysis on the HuBERT-base encoder fine-tuned for ASR on the VCTK corpus. The attack is evaluated on $\mathcal{D}_{f, \text{test}}$ to distinguish between inherent acoustic discriminability and identity memorization induced by training. As shown in Figure~\ref{fig:layer_probing}, both $\mathcal{M}_{\text{orig}}$ and $\mathcal{M}_{\text{gold}}$ exhibit high classification accuracy ($>0.8$) in early layers. This behavior is consistent with prior observations in SSL speech models, where early layer representations capture highly variant acoustic feature that naturally separate speakers regardless of training exposure.

A critical transition occurs at the 6th encoder layer. At this layer, the accuracy of $\mathcal{M}_{\text{gold}}$, which has never been exposed to the forget speakers, drops to near random chance of $0.52$.
In contrast, $\mathcal{M}_{\text{orig}}$ maintains a significantly higher accuracy of $0.75$.
Such a substantial performance gap between the original and gold models indicates that the separability cannot be solely explained by raw acoustic differences. Instead, it reveals that speaker-specific identity information of the forget speakers is encoded in this intermediate layer due to training exposure. 
While a notable performance gap persists across several subsequent layers, the difference between the two models eventually diminishes in the final layers as representations become increasingly specialized for linguistic transcription. Targeting the onset of such memorization, we select the 6th layer as the optimal intervention point for speaker-level unlearning.

\noindent \textbf{Layer-wise Effects of Unlearning} As shown in Figure~\ref{fig:layer_probing}, we analyze the layer-wise MIA accuracy of the unlearned model ($\mathcal{M}_{\text{SAID}}$) across the encoder hierarchy. 
In early encoder layers, $\mathcal{M}_{\text{SAID}}$ shows slightly lower accuracy than both $\mathcal{M}_{\text{orig}}$ and $\mathcal{M}_{\text{gold}}$, indicating that the unlearning process actively disrupts the acoustic signals associated with the forget speakers. This proactive erasure weakens identity-related signals in the early stages of representation, where identity information is typically most variant. The more informative comparison appears at the 6th layer, the selected intervention point. At this layer, $\mathcal{M}_{\text{SAID}}$ closely follows the behavior of $\mathcal{M}_{\text{gold}}$ and substantially reduces the gap previously observed between $\mathcal{M}_{\text{orig}}$ and $\mathcal{M}_{\text{gold}}$. This result indicates that the proposed identity dispersion effectively neutralizes the speaker-specific signatures associated with the forget speakers. While all models eventually converge toward random-chance accuracy in the deeper layers as they prioritize phonetic and linguistic content, $\mathcal{M}_{\text{unlearn}}$ ensures that residual identity signals are eliminated much earlier in the representation flow. This confirms that once the identity signatures are dissolved at the targeted layer, the model maintains its privacy state, effectively ensuring that no residual identity information remains within the encoder. 

\section{Conclusion}
In this work, we study speaker-level machine unlearning in ASR. While ASR models are designed to recover linguistic content independently of speaker identity, prior study~\cite{asr_memory} shows that speaker information still persist within their internal representations. Our layer-wise analysis confirms that SSL-based ASR models retain speaker-specific traces in their latent representations, creating potential privacy risks. Motivated by this, we explore a feature-level unlearning strategy that targets these identity structures directly in the latent space. Experimental results show that the influence of specific speakers can be removed while preserving general transcription capability, providing a promising direction toward privacy-aware speech recognition systems.
Beyond speaker identity, our framework can be extended to other feature-level unlearning tasks~\cite{speech_unlearning}, including lexical identity leakage where speaker-specific information may persist through characteristic vocabulary or phrasing patterns. 

\section{Generative AI Use Disclosure}
During the preparation of this manuscript, the author used ChatGPT to improve the language and readability of the text. After using this tool, the authors reviewed and edited the content as needed and take full responsibility for the final version of the paper. Tha AI tool was strictly limited to editing and polishing and was not used to generate a significant part of the original content of the manuscript.

\section{Acknowledgements}
This work was partly supported by the National Research Foundation of Korea(NRF) grant funded by the Korea government(MSIT) [No. RS-2025-24683892, 50\%], Institute of Information \& communications Technology Planning \& Evaluation (IITP) grant funded by the Korea government(MSIT) [No. RS-2022-II220320, 20220-00320, 45\%] and [No. RS-2021-II211343, 5\%]. The GPUs were supported by the Advanced GPU Utilization Support Program funded by the Ministry of Science and ICT and supervised by NIPA [No. 02-26-01-0285].

\bibliographystyle{IEEEtran}
\bibliography{references}

\end{document}